\documentclass[12pt]{article}
\usepackage{graphicx}
\usepackage[bookmarksnumbered,colorlinks,plainpages]{hyperref}

\begin{document}

\title{The $q$-nonadditivity of nonextensive statistics is not a true physical property}

\author{Q.A. Wang$^\dag$, C.J. Ou$^\ddag$, A. J.C. Chen$^\S$\\
{\it $^\dag$ISMANS, 44, Avenue F.A. Bartholdi, Le Mans, France}\\
{\it $^\ddag$College of Information Science and Engineering,}\\
{\it Huaqiao University, Quanzhou, China}\\
{\it $^\S$Department of Physics, Xiamen University, Xiamen, China}}

\date{}

\maketitle

\begin{abstract}
This is a note showing that, contrary to our lasting belief, the
nonadditivity $X(1\cup 2)=X(1)+X(2)+\alpha X(1)X(2)$ is not a true physical
property. $\alpha$ in this expression cannot be unique for a given system.
It unavoidably depends on how one mathematically divides the system and
cannot be used to characterize nonadditivity. As a matter of fact, its use
is mathematically inconsistent.
\end{abstract}

People who worked with $q$-nonextensive statistics\cite{Tsallis} are all
aware of a nonlinear relationship characterizing the nonadditivity of
entropy and energy which is an emblem of the theory and a central object of
polemics about the foundation of the theory (see for example \cite{Polemics}
and references there-in) and its relationship with thermodynamics (a brief
review can be found in \cite{Wang1}). The nonadditivity is given by, for a
total system composed of two subsystems 1 and 2:

\begin{eqnarray}                                            \label{1}
X(1\cup 2)=X(1)+X(2)+\alpha X(1)X(2)
\end{eqnarray}
where $X$ represents entropy, energy or other variables, $\alpha=1-q$ (or
$q-1$) (with positive $q$) is a real coupling constant between 1 and 2. For
the total system, as claimed in every statement relative to this equality,
$\alpha$ characterizes its nonadditivity. Hence for a given system and
situation, $\alpha$ is expected to be unique. This relationship has been
widely employed in the definition of temperature, pressure and other
intensive variables in order to relate the nonextensive statistical
mechanics to thermodynamics.

The origine of this nonadditivity was the reference \cite{Tsallis2} from the
hypothesis of product join probability $p(1+2)=p(1)p(2)$ in order to show
the nonadditivity of entropy $S$ given by $S(1\cup 2)=S(1)+S(2)+\alpha
S(1)S(2)$ which was subsequently used in the discussion of zeroth law of
thermodynamics with an additional nonadditivity of energy $U$ given by
$U(1\cup 2)=U(1)+U(2)+\alpha U(1)U(2)$ (see for example a historical review
in \cite{Wang1}.

We note here that this nonadditivity is false because, on the one hand,
$\alpha$ is necessarily dependent on the subsystems of a given system so
that it cannot be used for characterizing the (unique) nonadditivity of the
latter and, on the other hand, with $\alpha$ dependent on the subsystems, a
self-contradiction is unavoidable when one treats a system composed of more
than two subsystems.

Let us see a calculation for (homogeneous) gravitational systems\cite{Ou}
which shows that $\alpha_{1,2}$ indeed depends on the size of the two
subsystems. The gravitational potential energy $V$ of a solid spherical mass
of radius $r_2$ separated into a smaller inner sphere of radius $r_1<r_2$
and a spherical outer shell of thickness $r_2-r_1$ is non-extensive:
$V_{total}=V(1)+V(2)+\alpha_{1,2}V(1)V(2)$ where
$\alpha_{1,2}=(\frac{r_1^3r_2^2}{2}-\frac{r_1^5}{2})/(\frac{r_1^5r_2^5}{25}-
\frac{r_1^8r_2^2}{10}+\frac{3r_1^{10}}{50})$. It is obvious that this
(homogeneous) system does not have unique $\alpha$. Each way of partitioning
the system will result in a different value of $\alpha_{1,2}$ or $q$,
meaning that Eq.(\ref{1}) does not have any practical meaning for
characterizing the gravitation nonextensivity. It should be noticed that
here we look at only the potential energy. If we add the thermal energy
(heat) and the eventual kinetic energy of the global motion which are
normally extensive and additive, the nonadditivity of the total energy
cannot even take the form of Eq.(\ref{1}).

Then a question arises: can Eq.(\ref{1}) existe for certain if not all
homogeneous system with $\alpha$ independent of the subsystems? Let us give
an answer with entropy. For a given system divided into two subsystems 1 and
2, the total entropy is $S(1\cup 2)=S(1)+S(2)+\alpha S(1)S(2)$. We can write
$S(1)/S(2)=k_1/k_2$ where the positive ratio $k_1/k_2$ characterizes the
division. This means that we can also write $S(1)=k_1 x$ and $S(2)=k_2 x$
here $x$ is the common factor of $S(1)$ and $S(2)$. As a result, the total
entropy reads $S(1\cup 2)=(k_1+k_2)x+\alpha k_1k_2x^2$. Since $S(1\cup 2)$
is constant for a given system, we can let it be unity without lose of
generality. This means $(k_1+k_2)x+\alpha k_1k_2x^2=1$. This equation must
have positive solution for $x$ (entropy is positive). That is
$\Delta=(k_1+k_2)^2+4\alpha k_1k_2\geq 0$ and
$\alpha<\frac{\sqrt{\Delta}-(k_1+k_2)}{2k_1k_2}$. This last relationship
shows that $\alpha$ cannot be independent of $k_1$ and $k_2$, i.e., of the
decomposition.

One can reach a more precise result simply by saying that, since the total
system is given, Eq.(\ref{1}) must read $X(1)+X(2)+\alpha X(1)X(2)=C$ where
$C$ is a constant. This implies
$\alpha=\frac{C}{X(1)X(2)}-\frac{1}{X(2)}-\frac{1}{X(1)}$ so that $\alpha$
is in general dependent on $X(1)$ and $X(2)$. Unavoidably, each way of
partitioning a given system yields a different value of $\alpha$.

Now let us see a mathematical self-contradiction which happens in the
application of Eq.(\ref{1}). Suppose a total system partitioned into three
subsystems 1, 2 and 3, which is a frequently encountered case in physics and
chemistry. One can for example image three atomic clusters forming a bigger
one, or the above mass sphere partitioned into an inner sphere and two
exterior shells. Let the nonadditivity between the subsystems $i$ and $j$ be
denoted by $X(i\cup j)=X(i)+X(j)+\alpha_{i,j} X(i)X(j)\;\;\;i\neq
j=1,2\;or\;3$ where the constant $\alpha_{i,j}$ is the coupling constant of
the composite system $i+j$, i.e., between the subsystems $i$ and $j$. All
three $\alpha_{i,j}$ can be different. Now let us see the already
constructed total system with a given total $X(1\cup 2\cup 3)$ and imagine
to add $X(1)$, $X(2)$ and $X(3)$ in different order. We first add $X(1)$ and
$X(2)$ and then $X(3)$. The total $X(1\cup 2\cup 3)$ is given by
\begin{eqnarray}                                            \label{3}
X((1\cup 2)\cup 3)&=& X(1\cup 2)+X(3)+\alpha_{(1,2),3} X(1\cup 2)X(3)\\
&=&X(1)+X(2)+X(3)+\alpha_{1,2}X(1)X(2)\\ &+&              \nonumber
\alpha_{(1,2),3}[X(1)X(3)+X(2)X(3)]+\alpha_{1,2}\alpha_{(1,2),3}X(1)X(2)X(3)
\nonumber
\end{eqnarray}
where $\alpha_{(1,2),3}$ is the coupling constant between the composite
system $1\cup 2$ and the subsystem 3. However, if we first add $X(1)$ and
$X(3)$ and then $X(2)$, the result is
\begin{eqnarray}\label{4}
X((1\cup 3)\cup 2)&=&X(1)+X(2)+X(3)+\alpha_{1,3}X(1)X(3)\\ &+& \nonumber
\alpha_{(1,3),2}[X(1)X(2)+X(3)X(2)]+\alpha_{1,3}\alpha_{(1,3),2}X(1)X(2)X(3)
\nonumber
\end{eqnarray}
The above two operations must give the same result $X(1\cup 2\cup 3)$ since
the additions of different order are just mathematical operations or thought
experiences without physically disturbing the system. However, Eq.(\ref{3})
can be equivalent to Eq.(\ref{4}) only if all the coupling constants
involved are zero or equal. The first case (zero coupling) is for extensive
system. The second case is impossible as shown above.

The above analysis is simply mathematics. No doubt seems possible. One is
obliged to say that Eq.(\ref{1}) has no practical meaning in the description
of nonextensive system. It must be rejected, together with the hypothesis of
product join probability, from the nonextensive physics. The hypothetical
product join probability, being the origin of the nonadditivity Eq.(\ref{1})
for entropy and energy, is in fact unrealistic as a general rule for two
subsystems which are dependent on each other due to nonextensivity. This
hypothesis is in addition contradictory with probability theory. Although
one can say that there may be dependent subsystems having by chance the
product joint probability, such an accidental and unpredictable case must
not be used for formulating general theory.

In summary, we showed that the nonadditivity given by Eq.(\ref{1}) with
constant $\alpha$ is untrue. Its use may lead to contradiction in the
calculation of physical quantities for a composite system. We would like to
emphasize that this unexpected result and its mathematical logic do not
touch (as far as we see) the validity of the $q$-exponential probability
distribution and the nonextensive statistics based on the Tsallis entropy as
an axiom. However, since Eq.(\ref{1}) was the starting point for the
definition of intensive variables of nonextensive system, the already
established (though discussible) connection between the nonextensive
statistical mechanics and thermodynamics is definitely broken.

This is not an encouraging result, especially for the hard searching for the
thermodynamics describing the systems that do not satisfy the conventional
paradigms such as Gibbs-Shannon entropy formula, equilibrium condition,
thermodynamic limits, ergodicity, energy and entropy additivity and others.
Examples of these systems include finite system, long range interacting
systems, large fluctuating system and many complex systems such as scale
free network and social process to cite only some.

We thank Professor S. Abe for fruitful discussion.


\begin{thebibliography}{99}


\bibitem {Tsallis}
C. Tsallis, Introduction to Nonextensive Statistical Mechanics, Springer
2009

\bibitem {Polemics}
Special issue of {\em Comptes Rendus Physique} {\bf 7},(2006)

Special issue of {\em Europhysics news}, {\bf 36/6} (2005)

\bibitem {Wang1}
Q.A. Wang, Laurent Nivanen, Alain Le Mehaute, Michel Pezeril, {\em Europhys.
Lett.}, {\bf 65}(2004)606

\bibitem {Tsallis2}
C. Tsallis, {\em J. Stat. Phys.} {\bf 52}, 479(1988)

\bibitem {Ou}
C.J. Ou, et al, {\em Physica A} {\bf 387}, 5761(2008)

\end{thebibliography}
\end{document}